\documentclass[twocolumn,showpacs,preprintnumbers,amsmath,amssymb]{revtex4}
\ifx\pdfoutput\undefined
\usepackage[dvips]{graphicx}
\usepackage[dvipdfm,
            pdfstartview=FitH,
            CJKbookmarks=true,
            bookmarksnumbered=true,
            bookmarksopen=true,
            colorlinks=true, %注释掉此项则交叉引用为彩色边框(将colorlinks和pdfborder同时注释掉)
            pdfborder=001,   %注释掉此项则交叉引用为彩色边框
            citecolor=blue,%
            linkcolor=blue
            ]{hyperref}
\AtBeginDvi{} % GBK -> Unicode
\else
\usepackage[dvipdf,
            pdfstartview=FitH,
            CJKbookmarks=true,
            bookmarksnumbered=true,
            bookmarksopen=true,
            colorlinks=true,   %注释掉此项则交叉引用为彩色边框
            citecolor=blue,%
            linkcolor=blue       　
            ]{hyperref}
\usepackage[pdftex]{graphicx}
\fi

\usepackage{caption2}
\usepackage{graphicx}% Include figure files
\usepackage{dcolumn}% Align table columns on decimal point
\usepackage{bm}% bold math
\usepackage{subfigure}
\usepackage{subfigure}
\usepackage{picinpar}
\usepackage{picins}
\usepackage{epic}
\usepackage{caption2}
\begin{document}

%\preprint{}

\title{Formation of regular spatial patterns in ratio-dependent predator-prey model driven by spatial colored-noise}% Force line breaks with \\

\author{Quan-Xing Liu }
%\email{liuqx315@sina.com}%Lines break automatically or can be forced with \\
\author{Zhen Jin}%
 \email{jinzhn@263.net}
\affiliation{%
Department of mathematics, North University of China,\\
Taiyuan, Shan'xi, 030051, People's Republic of China
}%

\date{\today}% It is always \today, today,

\begin{abstract}
Results are reported concerning the formation of spatial patterns
in the two-species ratio-dependent predator-prey model driven by
spatial colored-noise. The results show that there is a critical
value with respect to the intensity of spatial noise for this
system when the parameters are in the Turing space, above which
the regular spatial patterns appear in two dimensions, but under
which there are not regular spatial patterns produced. In
particular, we investigate in two-dimensional space the formation
of regular spatial patterns with the spatial noise added in the
side and the center of the simulation domain, respectively.
\end{abstract}

\maketitle

\section{Introduction}

The influence of noise on nonlinear systems is the subject of
intense experimental and theoretical investigations. The most
well-known phenomenon is noise induces transition~\cite{WH1984} and
stochastic resonance~\cite{KW1995,HZH1998}, both showing the
possibility to transform noise into order.

In this paper, the problem of the predator-prey dynamical
system~\cite{CJ1999,FHI1993,YK1998,MB2005,MR2006} is revisited and
the results are presented concerning the noise aspects of the
formation of spatial patterns in two dimensions. To our knowledge,
the predator-prey dynamics in the absence of noise have been studied
extensively using mathematical models. A common feature of these
models is the prediction that the populations can cycle for some
parameter sets: i.e. the populations of predators and prey do not
settle to constant values, but rather, oscillate periodically in
time. There are considerable data from field studies and laboratory
experiments to support the existence of such population cycles. If a
small group of predators are introduced into an spatial uniform
population of prey, the predators will tend to invade the prey,
leaving behind a mixture of predators and prey. In Jonathan A.
Sherratt, Mark Lewis, Barry Eagan and Andrew Fowler's
works~\cite{JAS1997,JAS1995}, they have studied the behavior of such
invasions for cyclic populations. The results are somewhat
surprising: the invasion leaves behind spatiotemporal oscillations
falling into either periodic traveling waves or spatiotemporal
irregularity; mixed behavior also occurs.

One of the key issues in ecology is how environmental fluctuation
and species interaction determine variability of the population
density~\cite{JMG1999,SAL1997,JR2003}. The other topic is that
dynamic patterns in two-dimensional space have recently been
introduced into ecology~\cite{SN2002,SN2000,ZT2005,UD2000}.
Ecologists have mainly been interested in the dynamical consequences
of population interactions, often ignoring environmental variability
altogether in space. However, the essential role of environmental
fluctuations has recently been recognized in theoretical ecology.
Noise-induced effects on population dynamics have been subject to
intense theoretical
investigations~\cite{VJMG1998,AG1998,MA2000,MR2002,MR2004,MR2005,MR2006,HB2002,PC2003,NK2003}.
Moreover, ecological investigations suggest that population dynamics
is sensitive to noise. In spite of the obvious significance of this
circumstance, the role of non-equilibrium spatial fluctuations
(spatial noise) of environmental has not been investigated much in
the context of ecosystems.

Recently, the general N-species Lotka-Volterra and ratio-dependent
predator-prey dynamical systems are described as stochastic models
by Romi Mankin, Ako Sauga and et al. i.e. add the parameter noise
term in the classical models, where the carrying capacity of the
population is taken into account as dichotomous noise. For former
model, the result of the study has shown that the noise induces
discontinuous transitions from a stable state to an instable one in
Refs.~\cite{MR2002,MR2004}. The latter model~\cite{MR2006} has shown
that the phenomena of colored-noise-induced Hopf bifurcation and
corresponding reentrant transitions can appear in ecosystems. But
both models are studied in nonspatial. From the
Refs.~\cite{JAS1997,DA2002,MP1993} we know that the space plays a
crucial role in the ecosystem. In Ref.~\cite{MP1993}, the results
have shown that the Turing patterns occur in the ratio-dependent
predator-prey models. Thus, naturally arise a problem that how the
spatial noise influences the formation of spatial patterns or
species invasion in the ratio-dependent predator prey dynamical
systems? Here we address how the spatial noise affects the formation
of the spatial patterns on the coexistence equilibrium in the
reaction-diffusion predator-prey dynamical systems.

\section{Model}

\subsection{The ratio-dependent model with diffusion}
We consider the reaction-diffusion model for predator-prey
interactions. At any point $(x,y)$ at time $T$, the dynamics of
prey ($N(x,y;T)$) and predator ($P(x,y;T)$) populations are given
by a reaction-diffusion model with logistic growth of the prey and
a type II (or Michaelis-Menten-type) function response of the
predator
\begin{subequations}\label{eq:RD}
\begin{equation}\label{eq:RDa}
\frac{\partial N}{\partial
T}=rN(1-\frac{N}{K})-\frac{bN}{kP+N}P+D_{1}\nabla^{2}N,
\end{equation}
\vspace{-0.5cm}
\begin{equation}\label{eq:RDb}
  \frac{\partial P}{\partial T}=e\frac{bN}{kP+N}P-\mu P+D_{2}\nabla^{2}P,
\end{equation}
\end{subequations}
where $\nabla^{2}$ is the Laplacian operator in Cartesian
coordinates. The parameters $r$, $K$, $\mu$ and $e$ denote the
intrinsic growth rate, carrying capacity of the prey, the death
rate of the predator and the yield coefficient of prey to
predator, respectively. The constant $D_{1}$ and $D_{2}$ are prey
and predator diffusion coefficients, respectively.

The model \eqref{eq:RD} can be simplified by introducing the
dimensionless variables, $n=N/K$, $p=Pk/K$, $t=Tr$,
$x=x/\sqrt{D_{1}/r}$, and $y=y/\sqrt{D_{1}/r}$. The units of the
dimensional variables will be scaled. Thus, in the terms of these
dimensionless variables the model is simplified to
\begin{subequations}\label{eq:NewRD}
\begin{equation}\label{eq:NewRDa}
\frac{\partial n}{\partial t}=(1-n)n-\frac{\beta
np}{n+p}+\nabla^{2}n,
\end{equation}
\vspace{-0.5cm}
\begin{equation}\label{eq:NewRDb}
  \frac{\partial p}{\partial t}=\varepsilon\beta\frac{np}{n+p}-\eta p+d\nabla^{2}p,
\end{equation}
\end{subequations}
where four new dimensionless parameters naturally arise:
\begin{eqnarray}\label{eq:NewP}
  \beta=\frac{b}{kr}, \eta=\frac{\mu}{r},
  \varepsilon=ek, d=\frac{D_1}{D_2}.
\end{eqnarray}

From the Ref.~\cite{CJ1999}, we know that the ratio-dependent
predator-prey dynamical system \eqref{eq:NewRD} has a homogeneous
coexistence point, $(n_{s},p_{s})$, i.e. the coexistence point is
given by
\begin{eqnarray}\label{eq: CP}
  n_{s}=1-(\eta/\varepsilon)(\Delta),p_{s}=(\Delta-1)n_{s},
\end{eqnarray}
where $\Delta=\varepsilon\beta/\eta$. This coexistence point is
biologically meaningful ($n_{s}>0,p_{s}>0$) only if
$\eta>\varepsilon(\beta-1)$ and $\eta<\varepsilon\beta$

Since we focus on the spatial noise affects the formation of the
spatial patterns in the ratio-dependent predator-prey system, we
don't show the mathematical property here.

\subsection{Spatial noise in predator-prey model}

Environmental fluctuations and other factors (see the discussion)
are important components in an ecosystem. In
Refs.~\cite{RMM1972,RMM2001}, May have pointed out the fact that
because of environmental fluctuations, the carrying capacities,
death rates, birth rates, and other parameters involved with the
systems exhibit random fluctuations to a greater or lesser extent.
Because a multinoise fluctuation stochastic system is difficult to
consider analytically, it is generally reasonable to confine only
one (or two) parameter of the system to fluctuate. In our paper, a
random interaction with the environment (such as climate, disease,
etc) in some domain is taken into account by introducing a random
spatial noise ($\xi(x,y;t)$) to the prey. Henceforth, the dynamic
system \eqref{eq:NewRD} is used to numerical simulation. In
comparison with Refs.~\cite{MR2002,MR2004,MR2005,MR2006}, here the
difference is that we take into account the spatial noise in two
dimensions. For the spatial noise we choose the colored noise
$\xi(x,y;t)$, which accounts for density fluctuations at point
$(x,y)$. The Ornstein-Uhlenbeck process was used to calculate the
colored-noise (Appendix). To characterize the spatial noise, we
introduce parameter, $D$, the intensity of the noise. The
$\xi(x,y;t)$ has the following properties:
$$\langle \xi(x,y;t)\rangle =0$$ and
$$\langle\xi(x,y;t)\xi(x',y';t')\rangle=D^2\delta(x-x')\delta(y-y')e^{-|t-t'|/\tau}.$$
Obviously, $\tau$ is the noise correlation time. Although the
dynamical system~\eqref{eq:NewRD} absent of random spatial noise
term is well understood, there is little analytic theory for the
system in two-dimensional space. The equations are integrated
numerically with an implicit scheme using $200\times200$ an
$50\times200$ spatial grid sites, respectively. For nonlinear
equations, the discretization introduced by numerical methods may
generate spurious results. To test this possibility, the
resolution of the simulation result was increased with space and
time, and the system is integrated with a different scheme, a
finite difference method. In all case the same qualitative results
have been obtained. We assume zero flux boundary conditions for
all $t$ in the simulation.

\section{Results}

\begin{figure}
\includegraphics[width=4.2cm, height=4.2cm]{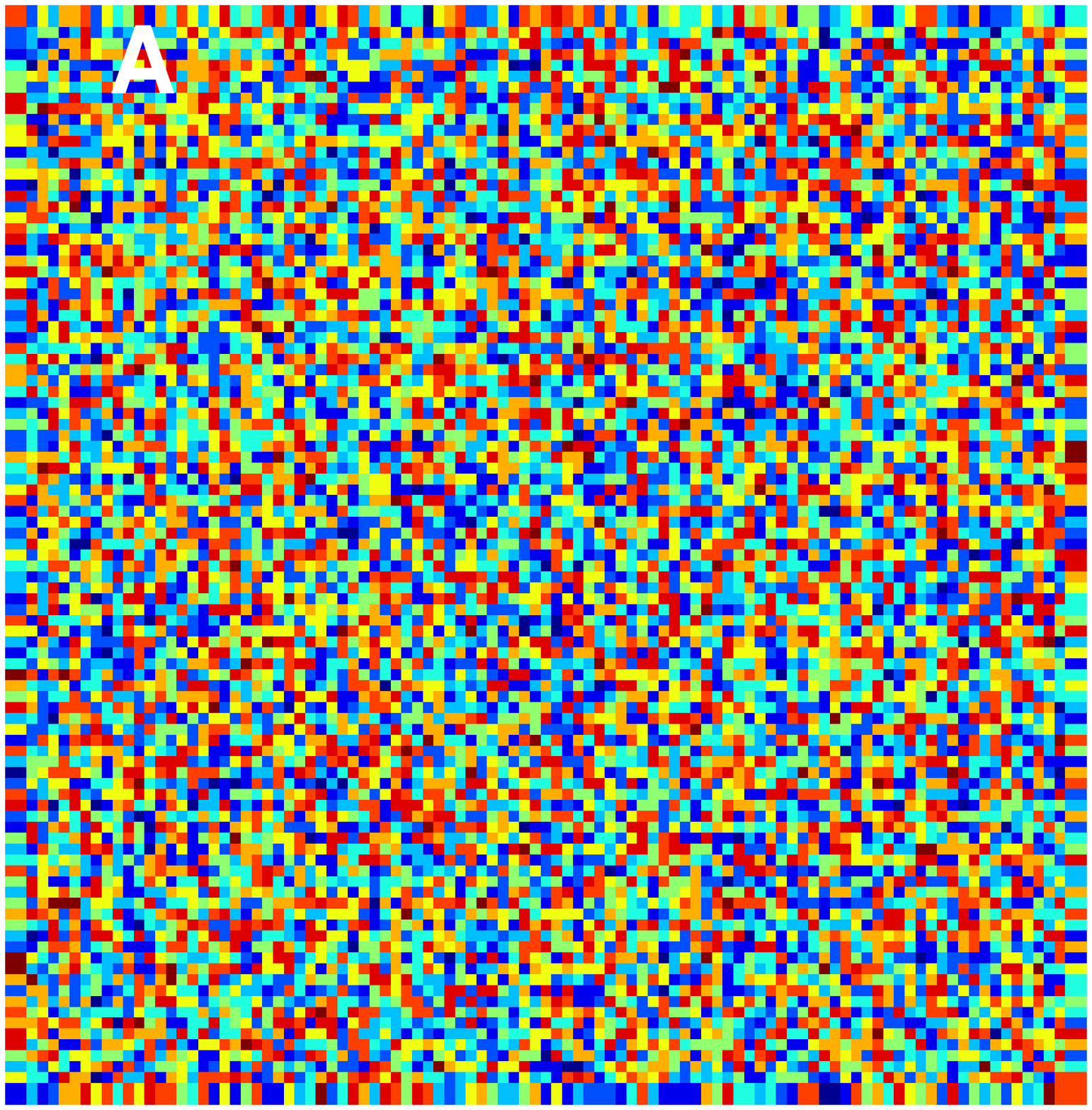}% Here is how to import EPS art
\includegraphics[width=4.2cm, height=4.2cm]{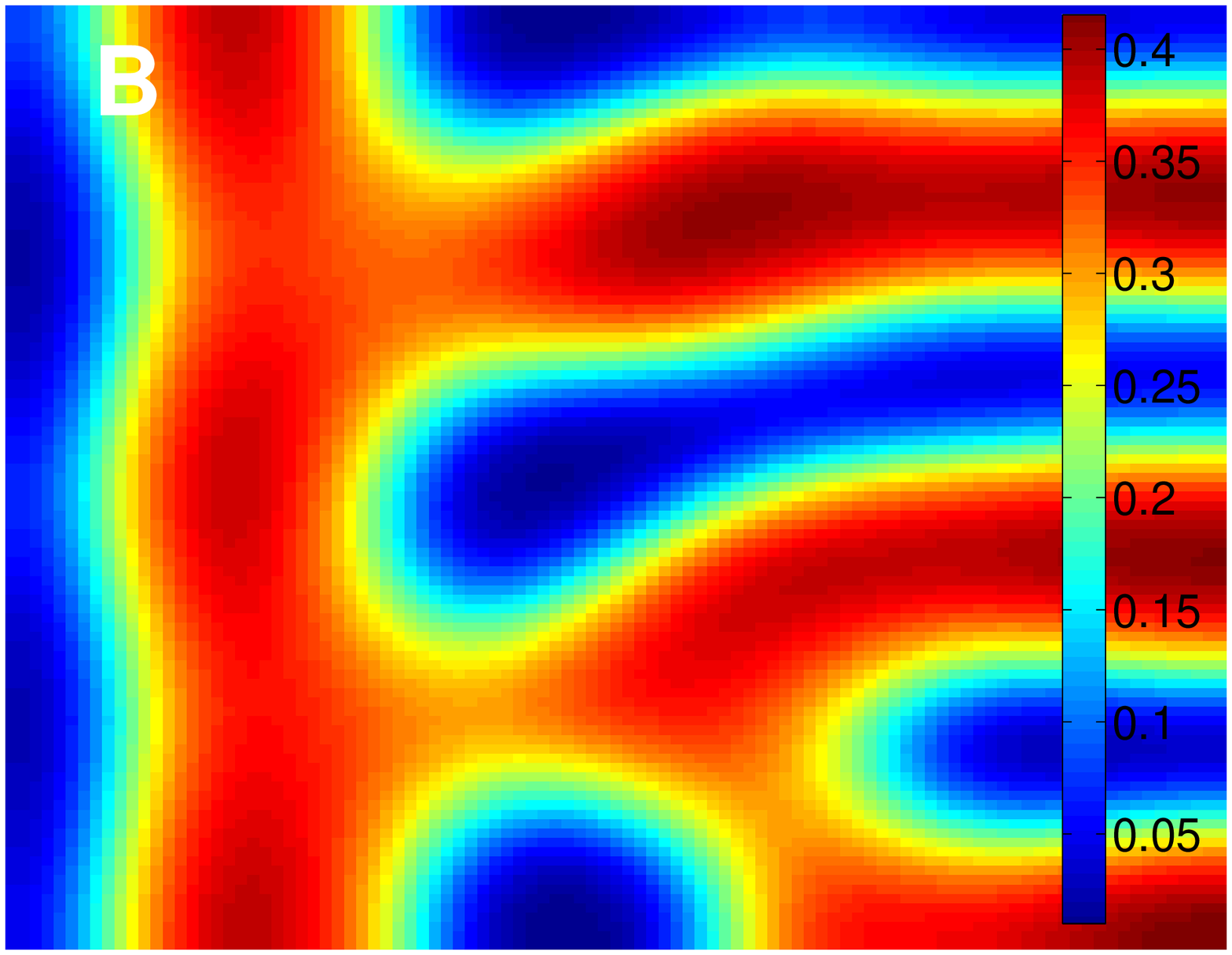}
\caption{\label{fig1}(Color online) Spatial patterns in two
dimensions absent of noise (in this case, the qualitative results
are the same as the Ref.~\cite{DA2002}). The numerical integration
of the dynamical system described by Eqs.~\eqref{eq:NewRD} has
been performed in a $200\times200$ lattices. (A) $t=0$ and (B)
$t=550$. The right inset shows the color corresponding to the
value of prey.}
\end{figure}

We study the effect of the spatial noise to the formation of the
spatial patterns. The results presented here are based on numerical
analysis for a set of parameters, $\beta=2$, $ \varepsilon=0.5$,
$\eta=0.6$, $d=10$ \cite{MISC1}, $\Delta h=0.5$ (space scales), and
$\Delta t=0.005$ (time step), chosen to obtain large-scale spatial
patterns in the absence of spatial noise (see Fig.~\ref{fig1}).
Moreover, from Turing Structures theory we know that these
parameters must be within the Turing space. In fact, the
authors~\cite{DA2002} have validated that these parameters are in
the Turing space.

Figure~\ref{fig1} illustrates the irregular spatial patterns
behavior of prey numbers in two dimensions absent of the spatial
noise. From this figure one can see that the spatial patterns are
irregular in two dimensions when they reach the
steady state.

%From this figure we are not difficult to discover that the system
%will appear periodic waves patterns in two dimensions.

\begin{figure}
\includegraphics[width=4.2cm, height=4.2cm]{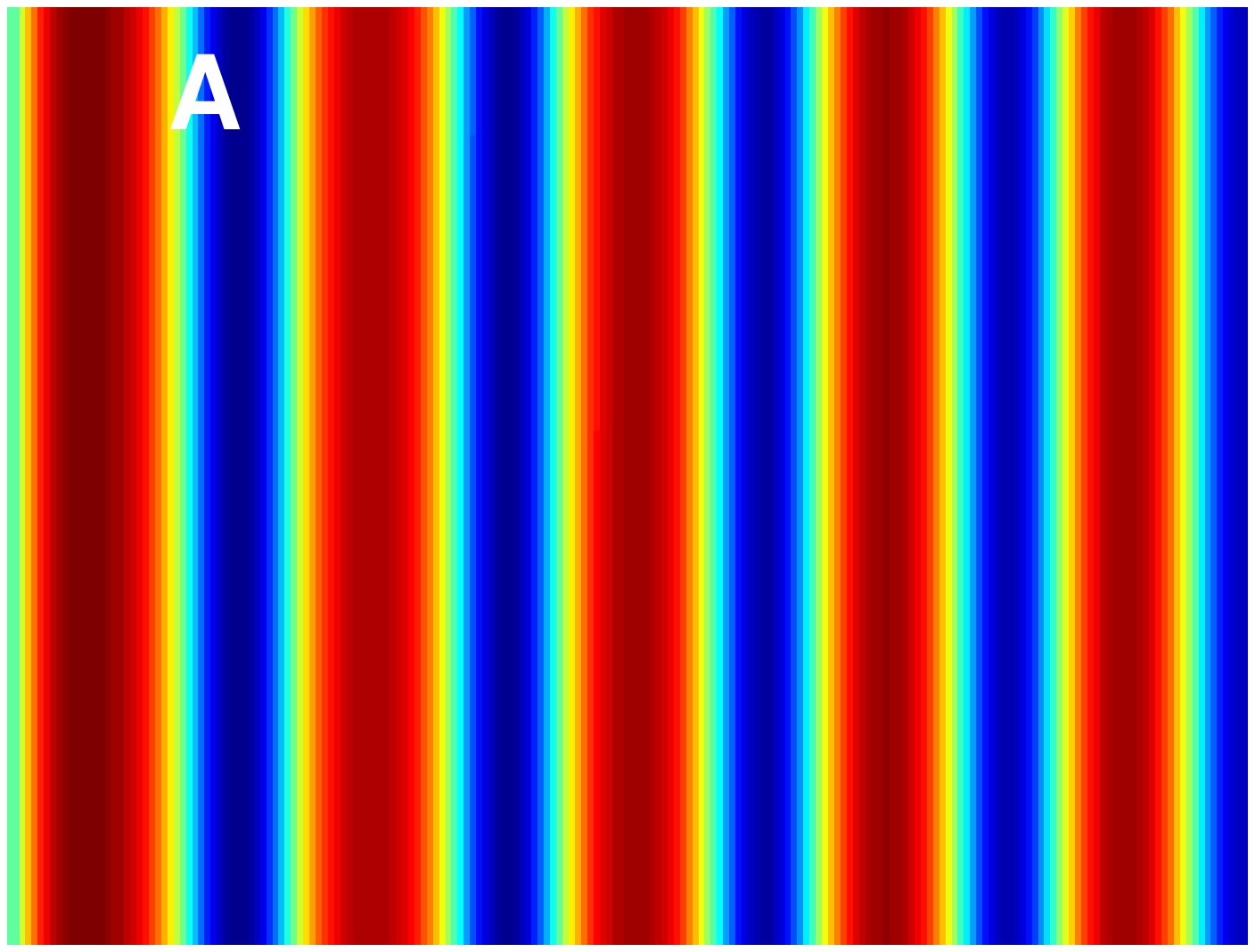}% Here is how to import EPS art
\includegraphics[width=4.2cm, height=4.2cm]{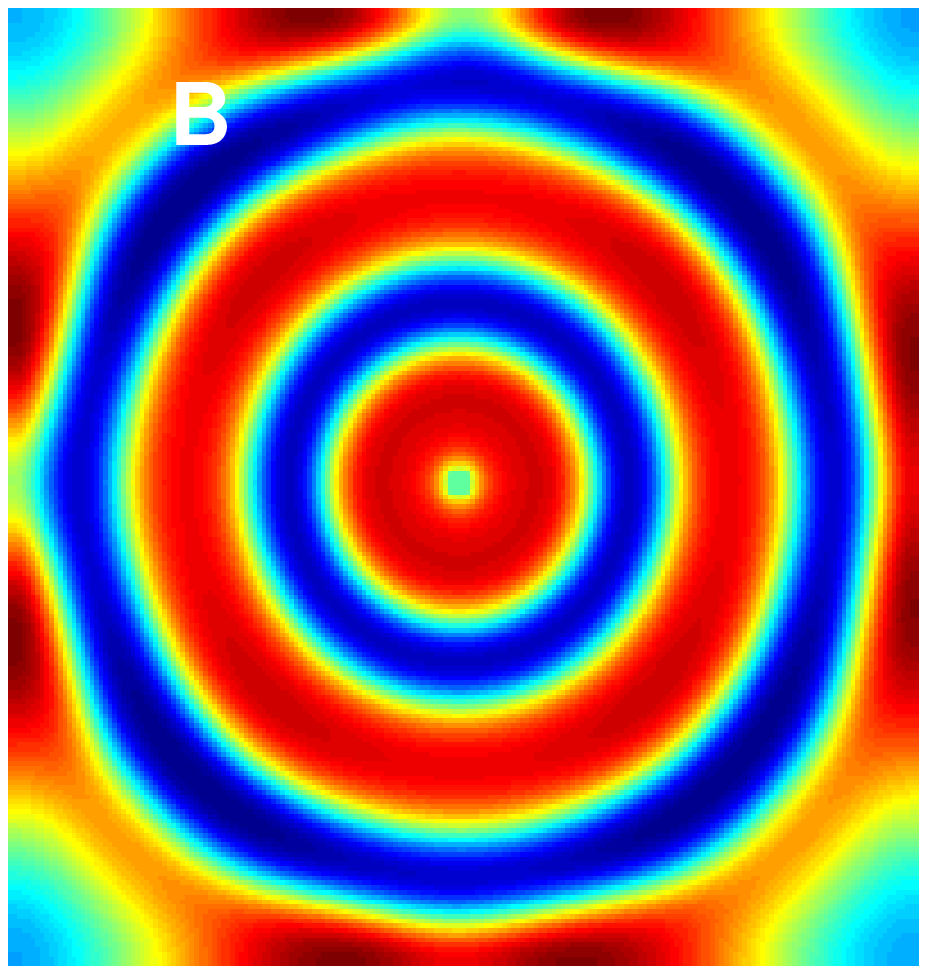}
\caption{\label{fig2}(Color online) Spatial patterns in two
dimensions driven by spatial noise. The numerical integration of
the dynamical system described by Eqs.~\eqref{eq:NewRD} has been
performed in $50\times200$ and $200\times200$ lattices,
respectively. (A) The spatial noise is added at the left-hand side
of the $50\times200$ lattice; (B) The spatial noise is added at
the center of the $200\times200$ lattices.}
\end{figure}

The main results in our paper show the effect of spatial noise with
respect to formation of the spatial patterns. In order to ensure the
facticity of the result, we have studied the system~\eqref{eq:NewRD}
in two cases driven by the spatial noise. First, we introduce the
spatial noise (here the spatial noise is perturbation on the
equilibrium value) at the left of the $50\times100$ spatial grid
sites. The domain of the spatial noise is set by $50\times 3$. The
ecological significance of the spatial noise can be explained for
the invasion of a prey population by predators in the space (see
Fig.~\ref{fig2}(A)). Second, we introduce the spatial noise (here
the spatial noise is the same as that in first) at the center of the
$200\times200$ spatial grid sites (see Fig.~\ref{fig2}(B)).

Figure~\ref{fig2} illustrates the formation of the regular spatial
patterns behavior of the prey numbers in two dimensions after the
spatial patterns reach the steady state, where the initially
condition (at $t=0$) add random and nonuniform small perturbations
to the equilibrium values. The maximum amplitude of the
perturbations is less than $0.0005$ and the noise correlation time
$\tau=1$. The intensity of the spatial noise is equal to 0.05.
Comparing Fig.~\ref{fig1} with Fig.~\ref{fig2}, one can easily find
that some characteristics are the same (here mainly relate to the
wavelength). But Fig.~\ref{fig2}(A) and Fig.~\ref{fig2}(B)
 both appear periodical structure patterns in two dimensions
 (called the spatial-period-2-structure). Additionally, these spatial
 patterns are regular in two dimensions.
 The only difference is that In Fig.~\ref{fig2}(A) the spatial patterns are
stripe but in Fig.~\ref{fig2}(B) target when the system reach steady state.

\begin{figure}
\includegraphics[width=4.2cm, height=4.2cm]{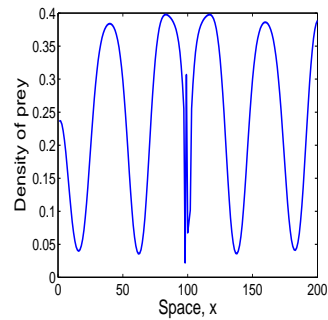}% Here is how to import EPS art
\caption{\label{fig3} This graph illustrates the regular patterns
driven by the spatial noise from other views. }
\end{figure}

If we change the direction of observing angle, the space vs density
plot can be seen more clearly. The space plot of the variable $n$ is
shown in Fig.~\ref{fig3}, where the horizontal axis represents the
horizontal spatial line crossing the center point. Fig.~\ref{fig3}
corresponds to Fig.~\ref{fig2}(B).

Finally, we increase the intensity of the spatial noise in the
system, and the results show that for every initial condition
there is a critical value of the intensity of noise, $D_{\rm c}$.
Here the critical value $D_{\rm c}\approx 0.02$ with foregoing
parameter. Above the critical value the regular spatial patterns
appear and maintain in two dimensions. However, under the critical
value, there doesn't exist regular spatial patterns even though
the simulation runs a long time. The qualitative results are the
same under different $D$ values (above the critical value), so we
only show typical results as in Fig.~\ref{fig2} when $D=0.05$.
Under the critical value, the result is not shown in our paper
since the spatial patterns are similar to Fig.~\ref{fig1}.

\section{ Conclusion and discussion}

In our paper we have considered the effect of the spatial noise
with respect to the spatial patterns in the ratio-dependent
predator-prey system in two dimensions. The results show that there
exits a critical value of the intensity of the spatial noise
($D_{\rm c}$) for the system~\eqref{eq:NewRD} when the parameters
are in the Turing space. Above the critical value the regular
spatial patterns appear in two-dimensional space, but under the critical value there
doesn't exit regular spatial patterns. Our key objective is that the
regular spatial patterns can appear in ratio-dependent predator-prey
model driven by spatial noise. Here, we use the intensity of the
noise to characterize the spatial noise, one can also use other
measures to characterize the spatial noise (e.g signal to noise
ratio).

Our results differ in several ways from the previous studies of
predator-prey equations in
ecology~\cite{JDM1989,MA2000,MR2002,MR2004,MR2005,MR2006}. Previous
studies have shown that the diffusion can induce instability and the
parameter noise can induce Hopf-bifurcation, discontinuous
transition and so on. But here we investigate the spatial noise in
the system~\eqref{eq:NewRD} in two dimensions. Our results suggest a
direction for further research, which is how to detect the spatial
fluctuation of the complex dynamics in natural environments. In
addition, two species predator-prey systems exhibit Turing spatial
patterns only under special conditions. In Ref.~\cite{DA2002}, the
study suggests that any strategy that involves interference between
predators should enhance patterns formation. Furthermore, the
process rates (prey birth rate, predator death rate, predator attack
rate) and the mobility of population size can also be determining
factors in the formation of biological patterns.

\section*{\label{app}Appendix}
The colored noise satisfies
\begin{eqnarray}\label{eq:app1}
  \langle\xi(t) \rangle  &=&0, \\
  \langle\xi(t)\xi(t') \rangle&=& D^{2}e^{-|t-t'|/\tau}. \nonumber
  \end{eqnarray}
It can be shown that the colored noise $\xi(t)$  my be calculated
form the following equation
\begin{eqnarray}\label{eq:app2}
  \frac{d\xi}{dt} &=& -\frac{1}{\tau}+\frac{D\sqrt{2\tau}}{\tau}
  \eta(t),
  \end{eqnarray}
where $\eta(t)$ is a Gaussian white noise source with
\begin{eqnarray}
  \langle\eta(t)\rangle &=& 0, \nonumber\\
  \langle\eta(t)\eta(t')
  \rangle &=& \delta(t-t').  \nonumber
\end{eqnarray}

Equation~\eqref{eq:app2} describes an Ornstein-Uhlenbeck
process~\cite{DJH2001}, and it can be shown that the  solution to
this equation satisfies Eqs. \eqref{eq:app1}.

\section{Acknowledgments}

This work was supported by the National Natural Science Foundation
of China under Grant No. 10471040 and the Science Foundation of
Shan'xi Province Grant No. 2006011009.

\newpage %Just because of unusual number of tables stacked at end
%\bibliography{English}% Produces the bibliography via BibTeX.

\end{document}